\documentclass[a4paper,12pt]{article}

\usepackage[utf8]{inputenc}   
\usepackage[T1]{fontenc}       

\usepackage[linesnumbered,ruled,vlined]{algorithm2e}
\usepackage{amsmath}
\usepackage{booktabs} 

\usepackage[dvips]{graphicx}
\usepackage{geometry}
\usepackage{verbatim}
\usepackage{amsfonts,amssymb}
\usepackage[normalem]{ulem}
\usepackage{color}
\usepackage{varioref,float,amsmath,amsthm}
\usepackage{mathtools}
\usepackage{multirow}

\usepackage{url} 

\usepackage{tcolorbox} 

\begin{document}

\begin{center}
.\\\vspace{7cm}
\noindent\textbf{\LARGE A Fractional Variational Approach to Spectral Filtering Using the Fourier Transform}\\ 

\vspace{2cm}
{\small\noindent{\bf Nelson H. T. Lemes}$^{1}$, {\bf José Claudinei Ferreira}$^{2}$ 
and {\bf Higor V. M. Ferreira}$^{1}$
}
\vspace{2cm}
\end{center}

\noindent
\begin{center}
{\small 
$^{1}$ Laboratory of Mathematical-Chemistry, Institute of Chemistry, Universidade Federal de Alfenas (UNIFAL), Alfenas, MG, Brazil  \\
$^{2}$ Institute of Exact Science, Universidade Federal de Alfenas (UNIFAL), Alfenas, MG, Brazil  \\
}
\end{center}
\vspace{\stretch{1}}
\noindent{$^*$

\newpage
\subsubsection*{Abstract}


The interference of fluorescence signals and noise remains a significant challenge in Raman spectrum analysis, often obscuring subtle spectral features that are critical for accurate analysis. 
Inspired by variational methods similar to those used in image denoising, our approach minimizes a functional involving fractional derivatives to balance noise suppression with the preservation of essential chemical features of the signal, such as peak position, intensity, and area. 
The original problem is reformulated in the frequency domain through the Fourier transform, making the implementation simple and fast. In this work, we discuss the theoretical framework, practical implementation, and the advantages and limitations of this method in the context of {simulated} Raman data, as well as in image processing. 
The main contribution of this article is the combination of a variational approach in the frequency domain, the use of fractional derivatives, and the optimization of the {regularization parameter and} derivative order through the concept of Shannon entropy. 
This work explores how the fractional order, combined with the regularization parameter, affects noise removal and preserves the essential features of the spectrum {and image}. 
Finally, the study shows that the combination of the proposed strategies produces an efficient, robust, and easily implementable filter.\\

\noindent
{\bf Keywords:} Fractional calculus; Variational methods; Fourier transform; Spectral filtering; Signal processing

\clearpage
\section{Introduction}

Raman spectroscopy is a powerful analytical technique used for molecular characterization in various fields, including chemistry, physics, and biomedical sciences. By measuring the inelastic scattering of light, it provides detailed information about molecular vibrations, allowing for the identification and characterization of substances. However, the presence of noise in Raman spectra can obscure subtle spectral features, making accurate data interpretation challenging.

In the case of Raman spectroscopy, noise is particularly concerning given that only one photon resulting from inelastic Raman scattering is detected for every $10^6$--$10^8$ incident photons, due to the low efficiency of the process involved.\cite{1} Another inherent difficulty is that some samples are fluorescent, which can severely compromise the results by masking the typically weak Raman peaks. In such cases, the spectrum may exhibit a ramp or hill of increasing intensity, generally covering the entire spectral range. Consequently, effective methods for fluorescence and noise removal are essential to improve the quality of Raman spectral data and enable accurate analysis.

Experimentally, the problem can be mitigated using special techniques such as SERS (Surface-Enhanced Raman Scattering), which amplifies the Raman signal in many cases by factors of $10^6$ up to $10^{14}$.\cite{2a,2b} In parallel, instrumentation has been improved to minimize optical losses in the spectrometer geometry and increase the quantum efficiency of the detector. After data collection, mathematical tools can help remove the fluorescent background and minimize noise. Therefore, signal processing becomes an essential step to achieve a reasonable signal-to-noise ratio (SNR).

The most common computational approaches for noise reduction in Raman spectra include the Savitzky–Golay filter, widely used due to its local polynomial fitting within a moving window, and low-pass filters, such as Fourier-based filters, which are effective in suppressing high-frequency fluctuations without compromising peak integrity.\cite{3b,3c} However, these methods can sometimes over-smooth the data, leading to a loss of important spectral features, or may be less effective against certain types of noise, such as the broad background often seen in fluorescence.

The variational method provides an elegant and powerful framework for noise reduction in signals and images, and it has been extensively employed for filtering and smoothing spectral data, including Raman spectroscopy. The central idea of the variational approach is to formulate the denoising problem as the minimization of a cost functional — a mathematical operator that maps a given function (in this case, the spectrum) from a functional space to the real numbers.

In general, this cost functional comprises two terms: the first enforces fidelity to the observed data, while the second imposes smoothness on the reconstructed signal. A regularization parameter controls the trade-off between these competing objectives.\cite{5a,5b,5c,5d,5e} Kalivas\cite{5f} and Tencate et al.\cite{5g} present a comprehensive review of the contemporary literature on the subject.

The Rudin–Osher–Fatemi (ROF) model, introduced in 1992, is a seminal mathematical framework for image and signal denoising that preserves edges and sharp features, making it particularly suitable for spectra such as Raman and fluorescence.\cite{rudin1992} The ROF model seeks a smooth function \( f(x) \) that closely approximates the noisy signal \( u(x) \) while suppressing abrupt variations. This is achieved by minimizing a variational functional that includes the total variation (TV) term \( |\nabla f| \). The inclusion of this term, controlled by a regularization parameter \( \lambda \), promotes smoothness while preserving significant features. Additionally, the functional includes a fidelity term that constrains the recovered signal not to deviate excessively from the original noisy data.

The ROF model serves as the foundation for numerous generalizations, in which the total variation term is modified or replaced. A modern generalization replaces the total variation \( |\nabla f| \) with the absolute value of the derivative of the recovered function \( \|\nabla f\|^2_2 \), leading to the classical Tikhonov functional.

In \cite{fractionalTikhonov}, the authors propose a filtering method based on Tikhonov regularization incorporating fractional derivatives, termed the Fractional Tikhonov Method (FTM). The objective was to remove random noise from experimental Raman spectra (specifically SERS data) without compromising critical spectral features. The method was applied to experimental SERS spectra of crystal violet dye in a colloidal dispersion of silver nanoparticles (AgNPs). This approach introduces an additional degree of freedom — the fractional order \( \alpha \) — allowing finer control over the smoothing process. However, it also increases computational complexity due to the use of fractional operators and the inversion of high-dimensional matrices.

In this context, the present work suggests an alternative strategy to minimize the functional by solving the smoothing problem efficiently in the frequency domain using the Fourier transform. This approach reduces computational costs while retaining the advantages of fractional regularization.

In this work, we present the theoretical framework, practical implementation, and the advantages and limitations of the proposed method in the context of both simulated and experimental Raman data, as well as in image processing. Furthermore, we investigate how the fractional order, together with the regularization parameter, influences noise suppression while preserving essential spectral features. The determination of the fractional derivative order and regularization parameter is performed automatically based on Shannon entropy.

This paper is organized as follows: Section 2 introduces the proposed variational model and its formulation in the frequency domain. Section 2 also presents the mathematical background of the generalized Fourier transform filter. Section 3 discusses the numerical implementation and the strategy for noise removal in one-dimensional simulated Raman spectra. A two-dimensional example is also presented in Section 3. Finally, Section 4 provides concluding remarks and potential extensions.

\section{Fundamentals}

The variational method is a mathematical technique that can be applied to transform the problem of minimizing an error function, and thus smoothing experimental data by removing noise, into a differential equation problem. When applied to a one-dimensional experimental curve or a two-dimensional surface in an imaging context, this method seeks to find a function that best fits the data by minimizing the differences between the observed data and a smoothed function, while penalizing abrupt changes and preserving essential characteristics such as edges or patterns in image processing.

\subsection{Filtering}

Let us begin with a simple denoising problem. Consider an experimental signal \( u(t) \), representing a one-dimensional curve. The objective is to obtain a smoothed version \( f(t) \) of this signal by minimizing the impact of the noise \( \eta(t) \) present in the observed data \( u(t) \):

\[
u(t) = f_0(t) + \eta(t).
\]

An effective filtering method should yield \( f(t) \approx f_0(t) \). Recovering the unknown signal \( f_0(t) \) is a typical ill-posed inverse problem, commonly addressed using Tikhonov regularization. In this framework, the functional \( J[f] \) to be minimized incorporates two competing terms: a fidelity term that ensures \( f(t) \) remains close to the observed data \( u(t) \), and a regularization term that imposes smoothness on \( f(t) \):

\[
J[f] = \|f - u\|^2 + \lambda \|f^\prime\|^2
= \int_{\mathbb{R}} |f(t) - u(t)|^2 dt + \lambda \int_{\mathbb{R}} | f^\prime(t) |^2 dt,
\]
where \(\| \cdot \|\) denotes the \(L^2\)-norm. The regularization parameter \(\lambda > 0\) controls the trade-off between fidelity to the data and the degree of smoothness imposed on the reconstructed signal.

The problem of filtering a noisy signal can be formulated as the minimization of this functional \(J[f]\), which is conveniently solved in the frequency domain. 

Defining the Fourier transform of an absolutely integrable function \(f(t)\) as

\[
\hat{f}(\omega) := \mathcal{F}\{f(t)\}(\omega) := \int_{\mathbb{R}} f(t) e^{-i\omega t} dt,
\]
it follows that, if \(\hat{f}\) is also absolutely integrable, then 

\[
f(t) = \frac{1}{2\pi} \int_{\mathbb{R}} \hat{f}(\omega) e^{i\omega t} d\omega,
\]
is a continuous function. Here we denote \(i^2 = -1\). Plancherel’s theorem \cite{Folland} ensures that the Fourier transform is well-defined for a function \(f \in L^2\), and the \(L^2\)-norm is preserved between the time and frequency domains, that is,

\[
\|f(t)\|^2 = \frac{1}{2\pi} \|\hat{f}(\omega)\|^2.
\]
Moreover, if \(f'\) is also in \(L^2\), it follows that

\[
i\omega \hat{f}(\omega) := \mathcal{F}\{f'(t)\}(\omega).
\]

Thus, the problem can be reformulated in the frequency domain using the Fourier transform:

\[
J[f] = \frac{1}{2\pi} \, \| \hat{f} - \hat{u} \|^2 + \frac{\lambda}{2\pi} \| \mathcal{F}\{f'(t)\} \|^2.
\]

For a cleaner representation, we adopt the convention in which 
the factor \(2\pi\) is absorbed into the regularization parameter and define the associated functional 

\begin{equation}
\label{def-E}
E[\hat{f}] = \| \hat{f} - \hat{u} \|^2 + \lambda \| \mathcal{F}\{f'(t)\} \|^2 = 
\int_{\mathbb{R}} |\hat{f} - \hat{u}|^2 d\omega
+ \lambda \int_{\mathbb{R}} |i\omega \hat{f}(\omega)|^2 d\omega
\end{equation}
\[
E[\hat{f}] =\int_{\mathbb{R}} |\hat{f}(\omega) - \hat{u}(\omega)|^2 + \lambda|i\omega \hat{f}(\omega)|^2 d\omega
\] 

Minimizing \(E[\hat{f}]\) with respect to \(\hat{f}(\omega)\), and using the Fréchet derivative on \(L^2\), leads to the first-order optimal  condition 
\[
\nabla E[\hat{f}] = 0 = \nabla J[f],
\]
or equivalently,

\[
\frac{d E[\hat{f}]}{d\hat{f}}(\omega) = -\hat{f}(\omega) + \hat{u}(\omega) + \lambda \omega^2 \hat{f}(\omega) = 0 \qquad \text{(almost everywhere)}.
\]
Note that $E[\hat{f}]$ is an integral whose kernel depends pointwise on $\omega$. Therefore, we can
minimize the integrand function independently for each $\omega$. Solving for \(\hat{f}\) yields the optimal filter result:

\[
\hat{f}^*(\omega) = \frac{\hat{u}(\omega)}{1 + \lambda \omega^2}.
\]
Finally, the solution in the original time domain is obtained by applying the inverse Fourier transform:

\[
f(t) = \mathcal{F}^{-1} \left[ \frac{\hat{u}(\omega)}{1 + \lambda \omega^2} \right]
= \mathcal{F}^{-1} \left[ \hat{h}(\omega) \cdot \hat{u}(\omega) \right].
\]

The transfer function 

\[
\hat{h}(\omega) = \frac{1}{1 + \lambda \omega^{2}}
\]
acts as a low-pass filter in the frequency domain, attenuating the high-frequency components of \(\hat{u}(\omega)\). The larger the parameter \(\lambda\), the stronger the smoothing effect, leading to a more pronounced suppression of high-frequency oscillations.

\subsubsection{Discrete Fourier Transform} 

When we have a square-integrable function \( f:\mathbb{R}\to \mathbb{R} \) that vanishes rapidly as \( t \to \pm \infty \), we can choose \( l>0 \) and write the equality

\[
f(x)=\frac{1}{2l}\sum_{n=-\infty}^{\infty} c_{n,l} \, e^{i n \pi x / l} 
= \frac{1}{2\pi}\sum_{n=-\infty}^{\infty} \frac{\pi}{l} c_{n,l} \, e^{i x \left(\frac{n \pi}{l}\right)}, 
\qquad x\in [-l,l],
\]
as the Fourier series representation of \( f \), with convergence in \( L^2([-l,l]) \) \cite{Folland}, where

\[
c_{n,l} = \int_{-l}^{l} f(x)\,e^{-i x \left(\frac{n \pi}{l}\right)}\,dx 
\approx \int_{-\infty}^{\infty} f(x)\,e^{-i x \left(\frac{n \pi}{l}\right)}\,dx
= \hat{f}\left(\frac{n \pi}{l}\right), 
\qquad n \in \mathbb{Z}.
\]

In this way, and keeping in mind the sampling theorem \cite{Folland}, we consider discrete versions of the previous results when implementing them in algorithms on a computer. Accordingly, we denote, for instance, the discrete Fourier transform as

\[
\hat{F}[k] = \Delta t \sum_{n=0}^{N-1} F[n] \, e^{-i 2\pi kn / N}, 
\quad\text{and}\quad 
F[n] = \frac{1}{N \Delta t} \sum_{k=0}^{N-1} \hat{F}[k] \, e^{i 2\pi kn / N},
\]
such that \( F[n] = f(n \Delta t) \) for \( n=0,\dots,N-1 \), \( \Delta t = 2l/N \), \( \omega = 2\pi/(N \Delta t) \), and \( \Delta \omega = 2\pi/(N \Delta t) \). 

In this study, the algorithm was implemented in MATLAB\footnote{Alternatively, the reader may use Python. Relevant guidance can be found at \url{https://docs.scipy.org/doc/scipy/tutorial/fft.html}.} using the \texttt{fft} and \texttt{ifft} routines. The \texttt{fft} function efficiently computes the discrete Fourier transform of a discrete signal, converting it from the time (or spatial) domain to the frequency domain. The \texttt{ifft} function computes the inverse Fourier transform, reconstructing the signal in the time (or spatial) domain from its frequency representation. The \texttt{fft} and \texttt{ifft} functions in MATLAB do not include the factor \( \Delta t \), but it can be included when needed. For the purposes of the present application, there is no need to include this constant.

\subsubsection{Two-dimensional filter}

The generalization of the previous discussion to two dimensions is straightforward. Considering that the functional $J$ is now defined as

\begin{eqnarray*}
J[f] &=& \|f-u\|^2 + \lambda \|\nabla f\|^2 \\
&=& \int\int_{\mathbb{R}^2} \left( |f(x,y) - u(x,y)|^2 + \lambda \left(\left(\frac{\partial f(x,y)}{\partial x}\right)^2 + \left(\frac{\partial f(x,y)}{\partial y}\right)^2 \right)\right) dx\,dy.
\end{eqnarray*}

Through the Fourier transform, we have the equality (\cite{Folland})

\[
\mathcal{F}\left(\frac{\partial f(x_1,x_2)}{\partial x_j}\right)(\omega_1,\omega_2) = i\omega_j \, \hat{f}(\omega_1,\omega_2), \qquad j=1,2,
\]
where the two-dimensional Fourier transform is defined by

\[
\hat{f}(\omega_1, \omega_2) = \iint_{\mathbb{R}^2} f(x, y) \, e^{-i 2\pi (\omega_1 x + \omega_2 y)} \, dx \, dy
\]
for \(f \in L^1(\mathbb{R}^2)\). Also, if \(\hat{f} \in L^1(\mathbb{R}^2)\), the inverse Fourier transform is defined by

\[
f(x,y) = \frac{1}{(2\pi)^2}\iint_{\mathbb{R}^2} \hat{f}(\omega_1, \omega_2) \, e^{i 2\pi (\omega_1 x + \omega_2 y)} \, d\omega_1 \, d\omega_2.
\]

Adopting the convention in which 
the factor $2\pi$ is absorbed into the {regularization} parameter, and letting $w = (w_1, w_2)$, we define the associated functional as

\begin{equation}
\label{def-Ec}
E[\hat{f}] = \| \hat{f} - \hat{u} \|^2 + \lambda \|\mathcal{F}\{\nabla f(x,y)\}\|^2 
= \| \hat{f} - \hat{u} \|^2 + \lambda \int_{\mathbb{R}^2} |i\omega \cdot \hat{f}(\omega)|^2 d\omega.
\end{equation}

Minimizing \(E[\hat{f}]\) with respect to \(\hat{f}(\omega)\), using the Fréchet derivative on $L^2$ as in the one-dimensional case, leads to the first-order optimal condition 
\[
\nabla E[\hat{f}] = 0 = \nabla J[f],
\]
or equivalently,

\[
\frac{d E[\hat{f}]}{d\hat{f}}(w) = -\hat{f}(w) + \hat{u}(w) + \lambda |\omega|^2 \hat{f}(w) = 0 \qquad \text{(almost everywhere)}.
\]
Solving for \(\hat{f}\) yields

\[
\hat{f}^*(\omega) = \frac{\hat{u}(\omega)}{1 + \lambda |\omega|^2},
\]
and it follows that 

\[
f(t) = \mathcal{F}^{-1} \left[ \hat{h}(\omega) \cdot \hat{u}(\omega) \right].
\]
The transfer function 

\[
\hat{h}(\omega) = \frac{1}{1 + \lambda |\omega|^{2}}
\]
acts as a low-pass filter in the frequency domain.

In the two-dimensional domain, 

\begin{equation}
\label{nabla}
\nabla J[f] = f - u - \lambda \Delta f,
\end{equation}
where $\Delta f$ is the Laplacian of $f$, and $-\Delta f = -\nabla(\nabla f) = (\nabla^* \nabla)f$, in which $\nabla^* = -\nabla$ is the adjoint of the gradient operator $\nabla$ on its domain. This implies that $-\Delta$ is a positive operator, and we can work with its square root $(-\Delta)^{\frac{1}{2}}$ in the norm equalities\footnote{The symbol
	$\Delta f =\frac{\partial^2 f(x,y)}{\partial x^2}+ \frac{\partial^2 f(x,y)}{\partial y^2}$ is the most common notation in mathematical analysis for the Laplacian of a function $f$, while in physics and chemistry the notation $\nabla^2 f$ 
	is usually employed, as it emphasizes the relation with the gradient operator.  The equality \eqref{nablab} follows from Green’s theorem, provided that $f$ is a $C^2$
	function with compact support; equivalently, $f$ is twice continuously differentiable and identically zero outside a ball of sufficiently large radius.}
\begin{equation}
\label{nablab}
\|(-\Delta)^{\frac{1}{2}}f\|^2 = \langle (-\Delta)^{\frac{1}{2}}f, (-\Delta)^{\frac{1}{2}}f \rangle = \langle f, -\Delta f \rangle = \langle f, (\nabla^* \nabla) f \rangle = \langle \nabla f, \nabla f \rangle = \|\nabla f\|^2,
\end{equation}
where \(\langle \cdot, \cdot \rangle\) denotes the inner product on $L^2$.

Here, the calculation of the Fourier transform and its inverse transform is performed using the functions \texttt{fft2} and \texttt{ifft2}.

\subsection{Fractional Filter}

The history of fractional calculus dates back to approximately 1695, when Leibniz first introduced the idea of a half-order derivative in his correspondence with L'Hôpital~\cite{ISBN:978-3030205232,DOI:10.1007/BFb0067095}. Since then, numerous definitions of the fractional derivative have been proposed. Among them, three are widely adopted in the literature: the Gr\"unwald--Letnikov, the Riemann--Liouville, and the Caputo formulations. In this work, we focus on the Riemann--Liouville approach, which is one of the most frequently used in both theoretical and applied contexts. For a comprehensive review of fractional calculus, see Podlubny~\cite{ISBN:9780125588409}. The choice of the Riemann--Liouville definition is motivated by its well-established Fourier transform properties, which are fundamental to the suggested  frequency-domain filtering approach.

The Riemann--Liouville fractional derivative is formally defined as~\cite{DOI:10.1111/j.1365-246X.1967.tb02303.x,ISBN:9780125588409}:

\[
[{_a}D{_t}^\alpha f](t) \coloneqq [D^nJ^{n-\alpha }f](t) = \frac{1}{\Gamma(n-\alpha)} \frac{d^n}{dt^n} \int_a^t \frac{f(u)}{(t-u)^{\alpha - n + 1}}\,du,
\]
where $\alpha$ is a real number satisfying $n-1 \leq \alpha \leq n$, with $n \in \mathbb{N}$, and $t > a > -\infty$ is the independent variable, in which $a$ is the lower bound. In this work, we adopt $a=-\infty$ and the Liouville fractional derivative, for which the Fourier transform is well defined.

The fractional operator is inherently nonlocal, as it depends on all values of $f(t)$ between the limits of the integral. This contrasts sharply with integer-order derivatives, which are local operators. An important property of the Caputo derivative is that the solution to the simplest fractional differential equation governing relaxation processes is given by the Mittag-Leffler function~\cite{DOI:10.1007/978-3-662-43930-2}. The Mittag-Leffler function is defined by a power series that depends on the parameter $\alpha$, serving as a {generalization} of the exponential function. In fact, it reduces to the exponential function when $\alpha = 1$. For $0 < \alpha < 1$, this function exhibits distinct decay rates for small and large times~\cite{DOI:10.21711/26755254/rmu202015}.

Some fundamental properties of the Riemann--Liouville fractional derivative are~\cite{DOI:10.1016/j.cnsns.2020.105338,DOI:10.1016/j.jcp.2019.03.008,DOI:10.1016/j.jcp.2014.07.019}:
\begin{itemize}
    \item[(a)] The Riemann--Liouville fractional derivative of order zero yields the original function.
    \item[(b)] The physical dimension of the fractional derivative of order $\alpha$ of a quantity $m$ with respect to time $t$ is given by the ratio of the unit of $m$ to the unit of $t$ raised to the power $\alpha$.
    \item[(c)] For integer $\alpha = n$, the Riemann--Liouville derivative coincides with the standard derivative of order $n$.
    \item[(d)] The Riemann--Liouville derivative of a constant is not zero. This property highlights the nonlocal nature of the operator, as it ``remembers" the history of the function.
\end{itemize}

Fractional derivatives have well-defined representations in the frequency domain. Specifically, the Fourier transform of the Liouville derivative of order $\alpha$ is given by

\[
\mathcal{F}\left\{ D_t^{\alpha} f(t) \right\}(\omega) = (i\omega)^{\alpha} \hat{f}(\omega),
\]
where $\hat{f}(\omega)$ denotes the Fourier transform of $f(t)$, and $\omega$ is the angular frequency.

In this work, we propose a novel spectral filtering method based on the fractional calculus of variations, formulated in the Fourier domain. A functional is defined to {penalize} deviations from the original noisy signal while simultaneously enforcing smoothness through a fractional-order derivative {regularization}:

\begin{equation}
\label{eq:funcional1D}
J[f] = \|f - u\|^2 + \lambda \|D^\alpha f\|^2 
= \int_{\mathbb{R}} |f(t) - u(t)|^2 \, dt + \lambda \int_{\mathbb{R}} |D^\alpha f(t)|^2 \, dt.
\end{equation}

The problem of filtering a noisy signal is thus formulated as the {minimization} of this functional. As in classical Tikhonov {regularization}, this can be conveniently solved in the frequency domain.

By applying the Fourier transform, as in Expression \eqref{def-E}, the equivalent functional becomes

\begin{equation}
\label{def-Eb}
E[\hat{f}] = \| \hat{f} - \hat{u} \|^2 + \lambda \|\mathcal{F}\{ D^\alpha f(t)\}\|^2 
= \| \hat{f} - \hat{u} \|^2 + \lambda \int_{\mathbb{R}} |i\omega^{\alpha} \hat{f}(\omega)|^2 \, d\omega,
\end{equation}
where $|i|^{2\alpha} = 1$. 
{Minimizing} this functional with respect to $\hat{f}(\omega)$ yields the solution in the frequency domain:

\[
\hat{f}(\omega) = \frac{\hat{u}(\omega)}{1 + \lambda \omega^{2\alpha}},
\]
and the corresponding solution in the time domain is obtained via the inverse Fourier transform:

\[
f(t) = \mathcal{F}^{-1} \left[ \frac{\hat{u}(\omega)}{1 + \lambda \omega^{2\alpha}} \right].
\]

The incorporation of the fractional derivative in the {regularization} term introduces an additional degree of flexibility, allowing the tuning of the filter's smoothness beyond what is achievable with classical integer-order derivatives. The parameters $\alpha$ and $\lambda$ control the degree of {regularization} and the smoothness properties. This is a key advantage of the fractional approach: it provides a continuous spectrum of smoothing effects, from minimal denoising ($\alpha \to 0$) to strong, classical Tikhonov-like smoothing ($\alpha = 1$). This flexibility is crucial for handling different noise characteristics and preserving subtle features.

Below, we present a pseudocode for the implementation of this method.\\

\begin{algorithm}[H]
\DontPrintSemicolon
\KwIn{Data file containing experimental data $u(t)$} 
\KwOut{Filtered curve $f(t)$}

\textbf{Parameters:} $N$ = number of data points; $\lambda$ =  regularisation parameter; $\alpha$ = fractional order \;

\textbf{Step 1: Read input data} \;
Read data file with pairs $(t_i, u(t_i))$ for $i = 1$ to $N$\;

\textbf{Step 2: Apply Fourier Transform} \;
Compute the centered Fourier transform: $\hat{u}(\omega) \gets \texttt{FFTShift}[\texttt{FFT}[u(t)]]$\;
Generate frequency vector $\omega_i$ corresponding to $t_i$\;

\textbf{Step 3: Construct fractional filter} \;
Define the spectral filter:
\[
\hat{h}(\omega) \gets \frac{1}{1 + \lambda |\omega|^{2\alpha}}
\]

\textbf{Step 4: Filter in frequency domain} \;
Apply filtering: $\hat{f}(\omega) \gets \hat{h}(\omega) \cdot \hat{u}(\omega)$\;

\textbf{Step 5: Inverse Fourier Transform} \;
Compute filtered signal: $f(t) \gets \texttt{IFFT}[\texttt{IFFTShift}[\hat{f}(\omega)]]$\;

\Return $f(t)$\;
\caption{Fractional Variational Filtering Using the Fourier Transform}
\end{algorithm}

\subsubsection{Two dimension version}
	
The generalization of the method to two dimensions is straightforward. Considering Expression \eqref{nablab}, the functional $J$ is now defined as

\[
J[f] = \|f - u\|^2 + \lambda \|(-\Delta)^{\alpha/2} f\|^2 
\]
\[
= \iint_{\mathbb{R}^2} |f(x,y) - u(x,y)|^2 \, dx \, dy + \lambda \iint_{\mathbb{R}^2} |(-\Delta)^{\alpha/2} f(x,y)|^2 \, dx \, dy,
\]
where $(-\Delta)^{\alpha/2}$ is the fractional Laplacian for some $\alpha > 0$.

The fractional Laplacian is a differential operator of non-integer order that {generalizes} the classical Laplacian to fractional orders. It is particularly useful for {modeling} non-local phenomena such as anomalous diffusion, where the change at a point depends on a larger {neighborhood} rather than just an infinitesimal one. This non-local property makes it a powerful tool for image processing, as it can preserve sharp edges while smoothing homogeneous regions more effectively than classical integer-order operators \cite{Riesz,Lischke}.

The most direct and operational definition of the fractional Laplacian is in the frequency domain, through the Fourier transform, in which case we have

\[
(-\Delta)^{\alpha/2} f(x,y) := \mathcal{F}^{-1}\left( |\omega|^\alpha \hat{f}(\omega) \right)(x,y),
\]
and, equivalently,
\[
\mathcal{F}\left\{(-\Delta)^{\alpha/2} f\right\}(\omega) = |\omega|^\alpha \, \hat{f}(\omega),
\]
where $1<\alpha\leq 2$ and
\[
|\omega| = \sqrt{\omega_1^2 + \omega_2^2}
\]
is the Euclidean norm of the frequency vector. For $\alpha = 2$, we have $(-\Delta)^{\alpha/2} = -\Delta$. This definition arises naturally in various contexts.\footnote{When the domain $\Omega$ is bounded and sufficiently regular, the eigenfunctions $\{\phi_k\}$ and the eigenvalues $\{\lambda_k\}$ of the Dirichlet Laplacian satisfy $\lambda_k>0$ and
	\[
	\begin{cases}
		-\Delta \phi_k = \lambda_k \phi_k, & \text{in }\Omega,\\[4pt]
		\phi_k = 0, & \text{on }\partial\Omega,
	\end{cases}
	\]
	where $\{\phi_k\}$ form an orthonormal set. This spectral decomposition allows one to define the fractional powers of the Laplacian by
	\[
	(-\Delta)^s u := \sum_k \lambda_k^s\, u_k\, \phi_k.
	\]
	
	In the specific case $\Omega = [0,1]^2$ the eigenfunctions are
	\[
	\phi_{m,n}(x,y) = \sin(m\pi x)\sin(n\pi y),
	\qquad m,n \in \mathbb{N},
	\]
	with corresponding eigenvalues
	\[
	\lambda_{m,n} = \pi^2 (m^2 + n^2).
	\]} However, this is neither the only nor necessarily the physically most appropriate way to define a fractional Laplacian \cite{Ortigueira}. The drawback of taking this as a starting point is that the Fourier transform is no longer available for bounded domains, although it remains applicable in this setting for zero boundary conditions.

The fractional Laplacian is preferred over the ``fractional gradient'' because the Laplacian is a scalar operator (acting from $\mathbb{R}^n$ to $\mathbb{R}$), while the gradient is a vector. When the Laplacian is raised to a fractional power, its scalar structure is preserved, which greatly facilitates both analysis and applications.  Moreover, the fractional Laplacian admits multiple characterizations on different domains, and there is currently no consensus in the literature regarding the most appropriate definition of the fractional Laplacian \cite{lischke,2022}.

By applying the Fourier transform, as in Expression \eqref{def-E}, the equivalent functional becomes

\begin{equation}
\label{def-Eb}
E[\hat{f}] = \| \hat{f} - \hat{u} \|^2 + \lambda \|\mathcal{F}\{ D^\alpha f(t)\}\|^2 = \| \hat{f} - \hat{u} \|^2 + \lambda \int_{\mathbb{R}^2} |\,|\omega|^{\alpha} \hat{f}(\omega)|^2 \, d\omega.
\end{equation}
\emph{Minimizing} this functional with respect to $\hat{f}(\omega)$ yields the solution in the frequency domain:
\[
\hat{f}(\omega) = \frac{\hat{u}(\omega)}{1 + \lambda |\omega|^{2\alpha}},
\]
and the corresponding solution in the spatial domain is obtained via the inverse Fourier transform:
\[
f(t) = \mathcal{F}^{-1} \left[ \frac{\hat{u}(\omega)}{1 + \lambda|\omega|^{2\alpha}} \right].
\]

This method demonstrates the scalability of the proposed approach, showing its applicability beyond one-dimensional signals to more complex data structures such as images.

The main feature that distinguishes it from the classical Laplacian is its non-locality. Whereas the classical Laplacian computes the ``curvature'' or ``Laplacian'' of a function at a point based solely on its infinitesimal neighborhood (making it a local operator), the fractional Laplacian takes into account the behavior of the function across its entire domain. This global dependence grants the operator the ability to capture the memory of the process. It is not limited to detect abrupt changes (such as peaks) or curvatures (as in a smooth signal) but also perceives the overall trend or long-range behavior of the signal.

\section{Results and Discussion}

To rigorously test the filter's performance under controlled conditions where the true signal and noise characteristics are precisely known, the Fractional Transform Method (FTM) was evaluated using synthetic data. This approach allows for an objective assessment of the filter's ability to recover the original signal. The synthetic data were generated with known and distinct peaks, along with an additive background noise component designed to mimic real fluorescence interference. This simulation enables the isolation and analysis of the filter's response to specific types of noise.

\subsection{One-dimensional Signal}

Raman spectroscopy is a powerful analytical technique that provides detailed information about the vibrational, rotational, and other low-frequency modes of molecules. When light, typically from a laser source, interacts with a molecule, the majority of photons are scattered elastically in a process known as Rayleigh scattering, in which their energy remains unchanged. A very small fraction of photons, however, undergo inelastic scattering, either gaining or losing energy by interacting with the vibrational modes of the molecule. This phenomenon is referred to as the Raman effect, and it forms the basis for Raman spectroscopy.
The resulting Raman spectrum displays the intensity of scattered light as a function of its frequency shift, typically expressed in wavenumbers ($\text{cm}^{-1}$) relative to the incident light. Each peak in the spectrum corresponds to a specific vibrational mode of the molecule, providing a molecular fingerprint that reflects its chemical structure and bonding environment \cite{xxx}.

\subsubsection{Simulated data}
\label{simulate}

The simulated data are intended to mimic the experimental data that contain an analytical signal and random noise. In this subsection, Raman spectra with Gaussian noise distributions were simulated and used to evaluate our method with regard to two different aspects: (i) the norm of the first-order derivative of the recovered signal, $\|{\bf f}^\prime\|$; and (ii) the norm of the difference between the analytical signal and the recovered signal, $\|{\hat{\bf f}}-{\bf f}\|$.

Therefore, the simulated Raman spectrum can be modeled as 

\[
\hat{\bf f} = {\bf f_0} + {\bf b} + {\bf n},
\]
where ${\bf b}$ is the baseline of the Raman spectrum ${\bf f_0}$ and ${\bf n}$ is the experimental noise. The pure signal ${\bf f_0}$ is the sum of pseudo-Voigt–shaped peaks \cite{Voight}, which have different positions, amplitudes, and widths. Each peak $k$ can be mathematically described as follows:

\[
f_k(\nu_i) = \eta_k L_k(\nu_i) + (1 - \eta_k) G_k(\nu_i),
\]
with $1 \leq i \leq 800$, where

\[
\begin{array}{ccc}
L_k(\nu_i) = \dfrac{\gamma_k}{\pi\left[(\nu_i - c_k)^2 + \gamma_k^2\right]}, & & 
G_k(\nu_i) = \dfrac{1}{\sqrt{2\pi}\gamma_k}\exp\left[-\dfrac{(\nu_i - c_k)^2}{2\gamma_k^2}\right].
\end{array}
\]

The simulated pure spectrum is then given by

\[
f_0(\nu_i) = \sum_{k=1}^M \alpha_k f_k(\nu_i),
\]
where $M = 2$ is the total number of peaks. The parameters $c_k$ are the peak positions, $\alpha_k$ are the maximum intensities, and $\gamma_k$ and $\eta_k$ are the shape control parameters. The parameters used in this study are $c_1 = 800$, $c_2 = 850$, $\alpha_1 = 0.8$, $\alpha_2 = 0.2$, $\gamma_1 = 4.9$, $\gamma_2 = 7.1$, $\eta_1 = 0$, and $\eta_2 = 0.5$. The baseline is ${\bf b} = 0$.

Finally, random noise was also added to the simulated spectra. The experimental noise is modeled using random numbers generated from a Gaussian distribution function with mean $\mu$ and standard deviation $\sigma$. The noise function ${\bf n}$ can be mathematically described as follows:

\[
N(\mu, \sigma) \rightarrow n(\nu_i),
\]
where $N$ denotes the distribution function from which $n(\nu_i)$ is obtained. The parameter that allows manipulation of the final noise level is the standard deviation $\sigma$, which in this study is set to $\sigma = 0.02$.   

\subsubsection{Shannon Entropy}

Shannon entropy is one of the central concepts of information theory, introduced by Claude E. Shannon in 1948 \cite{DOI:10.1002/j.1538-7305.1948.tb01338.x}. In this work, inspired by thermodynamics (Boltzmann entropy), he formulated an expression to measure how much information a signal carries, regardless of its semantic content.  
In the field of signal processing, Shannon entropy is used as a measure of the complexity or disorder of a signal.  

The Shannon entropy will be used to select the optimal parameters $\alpha$ and $\lambda$ in the functional $E[\hat{f}] = E^{\alpha,\lambda}[\hat{f}]$ (Equation \ref{def-Eb}). The central idea is that noisy signals tend to have higher entropy, as noise introduces greater randomness and uncertainty. Therefore, the strategy is to minimize the entropy of the filtered signal.  

Shannon entropy is defined as:
\[
H(P) = -\sum_i P_i \log_b(P_i),
\]
where $b$ is the base of the logarithm, and $P = \{P_1, P_2, \ldots, P_N\}$ represents a probability distribution.  

In the case of white noise, each amplitude value has an approximately equal probability of occurrence, that is, $P_i \approx 1/N$, which represents a condition of maximum uncertainty and, consequently, a high value $H(P) \lessapprox \log_b(N)$.  

In the present problem, the signal $f(t)$ does not directly constitute a probability distribution. For a Raman spectrum measured as pairs \(\{(\nu_i, I_i)\}\) (frequency/position \(\nu_i\) and intensity \(I_i \ge 0\)), the most straightforward way of defining \(P\) is  first replace  $I_i$ by $ I_i - \min{I}$ and  to normalize the intensities:
\begin{equation}
P_i = \frac{I_i^2}{\sum_{j=1}^N I_j^2}
.
\label{calculop}
\end{equation}

In this way, spectra with a few well-defined lines have low \(H\), whereas diffuse, noisy spectra or those rich in many peaks present higher \(H\).  

The filter used here produces an output signal \(\{(\nu_i, I_i)\}\) that depends on the parameters $\alpha$ and $\lambda$, denoted $I_i(\alpha, \lambda)$. This output generates the probability distribution $P_i(\alpha, \lambda)$. We also denote $H(\alpha, \lambda) = H(P_i(\alpha, \lambda))$. In this context, the pair $(\alpha, \lambda)$ that results in the lowest entropy value is considered the optimal set of parameters, i.e.,
\[
\{\alpha, \lambda\} = \underset{\alpha, \lambda}{\operatorname{min}} \, H(\alpha, \lambda).
\]

This optimization process seeks the smoothest possible signal that still retains its underlying structure, effectively finding the best balance between fidelity and regularization. This provides an objective criterion for parameter selection, which is often a significant challenge in filtering methods.

Another parameter that can be used to assess the image is the Tsallis entropy, which can more effectively capture long-range correlations, non-Gaussian behavior, and more complex textures. In the reference \cite{Tsallis}, the Tsallis entropy was computed from the two-dimensional histogram constructed using both the gray-level values of the pixels and their local average gray values. Reference \cite{Tsallis} demonstrates that, in terms of the quality of the thresholded images, Tsallis entropy yields better results than the Shannon method.

\subsubsection{Example}

In Figure~\ref{x2}, the minimum of $H$ defines the regions of $\alpha$ and $\lambda$ with the lowest noise levels.  
The figure also shows a broad region of low $H$ values, indicating that the method is robust and, therefore, only weakly sensitive to the parameters $\alpha$ and $\lambda$.  

As we can observe, the color map (blue for small $H$ and yellow for large $H$) displays a V-shape starting from $\alpha = 1$, showing that the region with low $H$ values increases as the value of $\alpha$ rises.  
This also suggests that different levels of filtering may lead to acceptable signals.  
Therefore, there are distinct pairs of values for $(\alpha, \lambda)$ that yield low values of $H$; however, this does not imply that the solutions are identical, as these combinations may lead to different results regarding oscillations, areas, and peak positions.  

From Equation~\ref{eq:funcional1D}, which defines the functional $J$, we know that a smaller $\lambda$ implies a greater weight on the difference between the obtained signal and the filtered signal, and a lower contribution from the penalty term. We therefore observe that increasing $\alpha$ allows smaller values of $\lambda$ to be considered.  

In Table~\ref{tab:resultados}, we find that, when keeping $\alpha$ fixed at $1$ and increasing $\lambda$, there is a decrease in $\|df\|$, that is, a reduction in oscillations. In this case, oscillations of both low and high frequency decrease, and consequently, the peak areas are affected. As can be seen in the table, as $\lambda$ decreases, the area of the first peak falls from 7 to 2.  

When $\alpha = 2.2$, we observe a different behavior. In this case, by increasing $\lambda$, $\|df\|$ decreases, though less markedly; however, the area under the curve at the first peak is preserved. This result was confirmed by keeping $\lambda = 10{,}000$ fixed and increasing the value of $\alpha$: we find that $\|df\|$ increases slightly, still remaining at acceptable levels (with low oscillation), and the value of the area is recovered as $\alpha$ increases from 1 to 3.4.  

The reference values for the area of the first peak, the position of the first peak (maximum), and $\|df\|$ are, respectively, 7.1311, 800.1252, and 0.0328.  

\clearpage
\begin{figure}[H]
   \centering
  \includegraphics[width=1\textwidth]{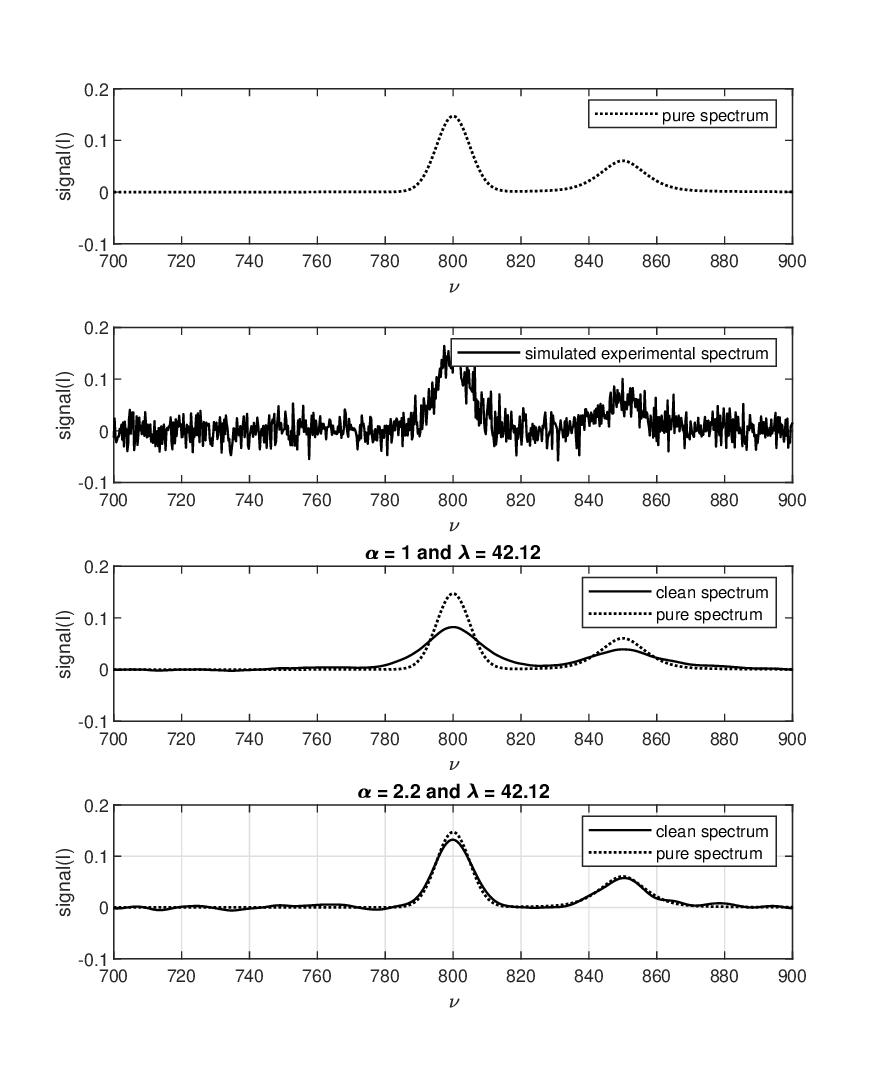}
  \vspace{-2.5cm}
  \caption{Filtered one-dimensional signal using the proposed fractional variational method.}
 \label{x1}
\end{figure}

\clearpage
\begin{figure}[H]
   \centering
  \includegraphics[width=.6\textwidth]{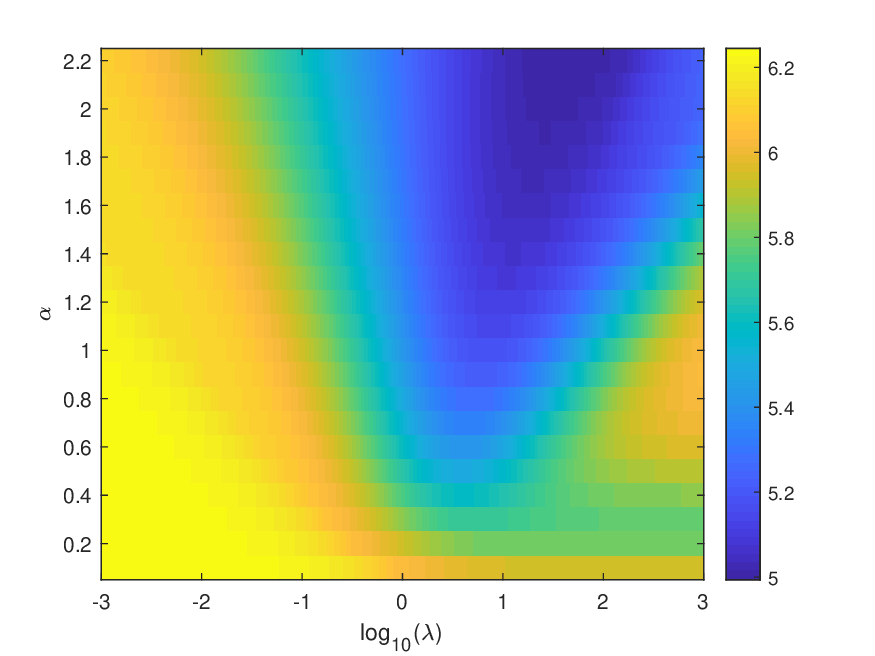}
  \caption{Optimization of parameters $\alpha$ and $\lambda$ based on the minimization of Shannon entropy.}
 \label{x2}
\end{figure}

\begin{table}[H]
\centering
\caption{Simulation results showing the influence of the fractional order $\alpha$ and the regularization parameter $\lambda$ on peak area, position, gradient norm $\|df\|$, and entropy $H$.}\vspace{0.5cm}
\begin{tabular}{cccccc}
\hline
{Peak area} & {Peak position} & $\|df\|$ & H & $\alpha$ & $\lambda$ \\
\hline
7.0897 & 800.38 & 0.47433   & 6.1263 & 1   & 0.01   \\
7.0859 & 800.63 & 0.15152   & 5.9461 & 1   & 0.1    \\
7.0813 & 800.63 & 0.043409  & 5.4328 & 1   & 1      \\
6.8139 & 800.13 & 0.022528  & 5.1822 & 1   & 10     \\
5.3728 & 800.13 & 0.0088216 & 5.6223 & 1   & 100    \\
3.2336 & 800.88 & 0.0021196 & 6.0272 & 1   & 1000   \\
2.0394 & 801.88 & 0.00033121& 6.0827 & 1   & 10000  \\
\hline
7.0770 & 800.88 & 0.10396   & 5.9475 & 2.2 & 0.01   \\
7.0928 & 800.88 & 0.054064  & 5.6248 & 2.2 & 0.1    \\
7.1188 & 800.13 & 0.036717  & 5.2815 & 2.2 & 1      \\
7.1377 & 799.87 & 0.031688  & 5.0382 & 2.2 & 10     \\
7.1488 & 799.87 & 0.026224  & 5.0071 & 2.2 & 100    \\
6.9749 & 800.13 & 0.018863  & 5.2330 & 2.2 & 1000   \\
6.0458 & 799.87 & 0.010218  & 5.4848 & 2.2 & 10000  \\
\hline
2.0394 & 801.88 & 0.00033121& 6.0827 & 1.0 & 10000  \\
3.4065 & 802.63 & 0.0021671 & 6.0817 & 1.4 & 10000  \\
4.9190 & 800.63 & 0.0049247 & 5.9069 & 1.8 & 10000  \\
6.0458 & 799.87 & 0.010218  & 5.4848 & 2.2 & 10000  \\
6.8062 & 800.13 & 0.016064  & 5.3496 & 2.6 & 10000  \\
7.1178 & 800.13 & 0.020082  & 5.3361 & 3.0 & 10000  \\
7.1881 & 800.13 & 0.022970  & 5.2262 & 3.4 & 10000  \\
\hline
\end{tabular}
\label{tab:resultados}
\end{table}

\clearpage
\subsection{Two-dimensional signal}

To apply the proposed procedure to color images, the method must be extended to handle each color channel independently. Color images in the RGB (Red, Green, Blue) format are represented by three pixel $M\times N$ matrices, one for each color channel.  
The input color image can be viewed as a function $u(x, y)$ and is decomposed into three separate channels:  
Red channel ($u_R$),  
Green channel ($u_G$), and  
Blue channel ($u_B$).  
Each of these channels is a grayscale image containing the intensity information for that color.

The method of minimizing the energy functional, which involves the Fourier transform and the fractional Laplacian, is applied to each channel individually. This means that the Fourier transform (and its discrete version) is computed for each channel:  

\[
\hat{u}_R(\omega), \quad \hat{u}_G(\omega), \quad \hat{u}_B(\omega),
\]
where $\omega=(\omega_1, \omega_2)$.

Then, the minimization equation  

\[
\hat{f}(\omega) = \frac{\hat{u}(\omega)}{1 + \lambda |\omega|^{2\alpha}},
\]
is applied to each of them:

\begin{align*}
\hat{f}_R(\omega) &= \frac{\hat{u}_R(\omega)}{1 + \lambda |\omega|^{2\alpha}} \\
\hat{f}_G(\omega) &= \frac{\hat{u}_G(\omega)}{1 + \lambda |\omega|^{2\alpha}} \\
\hat{f}_B(\omega) &= \frac{\hat{u}_B(\omega)}{1 + \lambda |\omega|^{2\alpha}}
\end{align*}

The filtered solutions in the frequency domain, $\hat{f}_R$, $\hat{f}_G$, and $\hat{f}_B$, are then transformed back into the spatial domain using the inverse Fourier transform ($\mathcal{F}^{-1}$):

\begin{align*}
f_R(x,y) &= \mathcal{F}^{-1}\{\hat{f}_R(\omega)\} \\
f_G(x,y) &= \mathcal{F}^{-1}\{\hat{f}_G(\omega)\} \\
f_B(x,y) &= \mathcal{F}^{-1}\{\hat{f}_B(\omega)\}
\end{align*}

Finally, the three processed channels, $f_R(x,y)$, $f_G(x,y)$, and $f_B(x,y)$, are combined to form the final color image, $f(x,y)$.  

This approach, which treats each color channel independently, is a standard method in most image-processing algorithms, as it assumes that the most relevant spatial and frequency correlations for processing occur within each channel $R$, $G$, and $B$. We calculate $H(P)$ for each channel:

\[
H = -\sum(P  \log_2(P))
\]
where $P$ is given by

\[
P_{k} = \frac{I_{k}}{\sum_{k}^{M\times N} I_{k}}
\]

In color images, the total Shannon entropy $H$ is calculated separately for each channel $R$, $G$, and $B$, and

\[
H = H_R + H_G + H_B
\]

For a color image $I$ with $R$, $G$, and $B$ channels, each channel is treated as an independent intensity image.

Shannon entropy in images quantifies the amount of information contained in pixel intensities. When entropy is high, the intensity distribution tends to be more uniform, indicating that many gray levels or colors are present, which generally corresponds to images with greater richness of detail, texture, or, in some cases, higher noise levels. Conversely, low entropy indicates that most pixels are concentrated in a few intensity levels, characterizing homogeneous or smoothed regions with little contrast variation. Thus, $H$ provides an objective measure of the “active information” in the image, reflecting its visual complexity and the diversity of details present. An improvement in post-filtering $H$ indicates effective noise suppression.

\subsection{Examples}

The noise added to the image is of the additive Gaussian type, characterized by a normal distribution with zero mean and a variance of 0.01, which implies a standard deviation of 0.1. This noise manifests visually as a granular pattern of random intensities superimposed on the original image, introducing stochastic fluctuations in all pixels. Regarding quality metrics, the noise artificially increases contrast (measured via the standard deviation of the pixels), as it amplifies the dispersion of luminous intensities. Simultaneously, it drastically raises the apparent sharpness (calculated by the gradient magnitude), since the abrupt variations of the noise simulate false edges and details. Finally, Shannon entropy also increases because the noise adds randomness and reduces statistical redundancy, expanding the diversity of pixel values and making the image histogram more complex. Although these metrics suggest numerical improvements, in practice they represent degradation of visual quality, since the noise masks genuine details and introduces unwanted artifacts.

Three quantities were employed to characterize the image: contrast ($\sigma$), 
sharpness ($\|\nabla f\|^2$), and Shannon entropy ($H$).

Contrast measures the difference between light and dark areas.  
High contrast aids in distinguishing objects and details.  
It is calculated as the standard deviation of all pixels:

\[
\sigma = \sqrt{\frac{1}{M\times N} \sum_{i=1}^{M}\sum_{j=1}^{N} (I_{i,j} - \mu)^2}
\]
where \( I_{i,j} \) is the intensity value of each pixel (0–255) for each channel, and $\mu$ is the mean value.  
However, excessively high contrast may saturate regions, amplify noise, and lead to loss of detail. In such cases, spots and imperfections become more visible.

Another parameter used is sharpness, which aims to assess the clarity of details and the definition of light-to-dark transitions.  
To evaluate sharpness, we use the gradient, which measures the rate of change of pixel intensity. A high gradient value indicates abrupt edges, while a low gradient corresponds to smooth transitions and uniform areas. We denote \( G_x(i,j) \) as an approximation of the partial derivative of $u(x,y)$, that is:

\[
G_x(i,j) = I(i,j+1) - I(i,j) \approx  \frac{\partial I}{\partial x}.
\]
Finally, we compute the average:

\[
\|\nabla I\|^2 \approx \frac{1}{MN} \sum_{i=1}^{M} \sum_{j=1}^{N} \sqrt{G_x(i,j)^2 + G_y(i,j)^2},
\]
when we view the matrix $I$ as a function $I(x,y)$, which is null when $(x,y)$ lies outside a rectangle in $\mathbb{R}^2$.

Analysis of the results presented in the table demonstrates how the filter parameters \( \alpha \) and \( \lambda \) directly affect image contrast, sharpness, and Shannon entropy \( H \).  
The reference image values are 0.2598, 7.8226, and 0.0976 for contrast, sharpness, and Shannon entropy, respectively.

In Table~\ref{tabela2}, it can be observed that for \( \alpha = 1.0 \), as \( \lambda \) increases from \( 10^{-2} \) to \( 10^{4} \), both contrast and sharpness decrease.  
Contrast declines from \( 2.57 \times 10^{-1} \) to \( 1.32 \times 10^{-1} \), while sharpness falls sharply from \( 9.29 \times 10^{-2} \) to just \( 6.86 \times 10^{-4} \).  
Shannon entropy also decreases (from 7.86 to 6.89), suggesting that increasing \( \lambda \) acts as a regularization mechanism that smooths the image, making it more homogeneous and reducing intensity diversity.  
The same behavior is observed for \( \alpha = 2.2 \), although in this case entropy remains more stable around 7.5–7.8, indicating greater robustness to smoothing.

In contrast, when fixing \( \lambda \) (e.g., \( \lambda = 10^{4} \)) and varying \( \alpha \) between 1.0 and 3.4, the effect is reversed:  
contrast gradually increases from \( 1.32 \times 10^{-1} \) to \( 2.23 \times 10^{-1} \), and sharpness rises from \( 6.85 \times 10^{-4} \) to \( 7.79 \times 10^{-3} \).  
Shannon entropy also increases slightly (from 6.89 to 7.57), proposing that the image recovers structural complexity and intensity diversity as \( \alpha \) increases.  
Thus, the parameter \( \alpha \) acts as an enhancer of image details, promoting edge sharpening and increasing perceived information.

Critically, it can be stated that \( \lambda \) serves as a smoothing agent: very high values drastically reduce contrast, sharpness, and entropy. Excessively high values result in blurred and uninformative images.  
On the other hand, the parameter \( \alpha \) plays an opposite role, preserving or even enhancing details, increasing entropy, and promoting greater structural distinction.  
The choice of parameters should therefore balance these opposing effects: high values of \( \alpha \) combined with low \( \lambda \) may amplify image noise, while high \( \lambda \) and low \( \alpha \) may lead to significant loss of information.

\clearpage
\begin{figure}[H]
   \centering
  \includegraphics[width=1\textwidth]{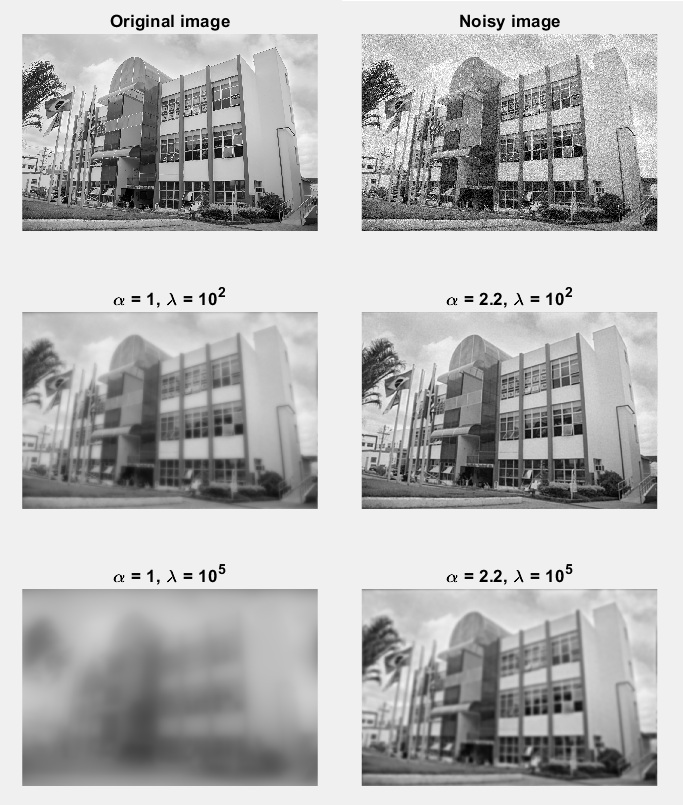}
  \caption{Application of the fractional variational filter to a 2D image.}
 \label{x}
\end{figure}

\clearpage
\begin{table}[H]
\centering
\begin{tabular}{ccccc}
\toprule
Contrast ($\sigma$) & Sharpness ($\|\nabla f\|$) & $H$ & $\alpha$ & $\lambda$ \\
\midrule
$2.5747 \times 10^{-1}$ & $9.2949 \times 10^{-2}$ & $7.8617$ & $1.0$ & $ 10^{-2}$ \\
$2.4745 \times 10^{-1}$ & $6.9288 \times 10^{-2}$ & $7.7959$ & $1.0$ & $ 10^{-1}$ \\
$2.3372 \times 10^{-1}$ & $2.9942 \times 10^{-2}$ & $7.6259$ & $1.0$ & $ 1$ \\
$2.2088 \times 10^{-1}$ & $9.8039 \times 10^{-3}$ & $7.5495$ & $1.0$ & $ 10$ \\
$2.0440 \times 10^{-1}$ & $3.7956 \times 10^{-3}$ & $7.4747$ & $1.0$ & $ 100$ \\
$1.7935 \times 10^{-1}$ & $1.5970 \times 10^{-3}$ & $7.3234$ & $1.0$ & $ 10^{3}$ \\
$1.3235 \times 10^{-1}$ & $6.8606 \times 10^{-4}$ & $6.8975$ & $1.0$ & $ 10^{4}$ \\
\midrule
$2.4961 \times 10^{-1}$ & $7.3874 \times 10^{-2}$ & $7.8133$ & $2.2$ & $10^{-2}$ \\
$2.4229 \times 10^{-1}$ & $4.7483 \times 10^{-2}$ & $7.7106$ & $2.2$ & $10^{-1}$ \\
$2.3653 \times 10^{-1}$ & $2.7488 \times 10^{-2}$ & $7.6342$ & $2.2$ & $1$ \\
$2.3101 \times 10^{-1}$ & $1.5492 \times 10^{-2}$ & $7.5904$ & $2.2$ & $10$ \\
$2.2572 \times 10^{-1}$ & $9.5345 \times 10^{-3}$ & $7.5699$ & $2.2$ & $100$ \\
$2.1989 \times 10^{-1}$ & $6.3762 \times 10^{-3}$ & $7.5553$ & $2.2$ & $10^{3}$ \\
$2.1310 \times 10^{-1}$ & $4.4033 \times 10^{-3}$ & $7.5355$ & $2.2$ & $10^{4}$ \\
\midrule
$1.3229 \times 10^{-1}$ & $6.8572 \times 10^{-4}$ & $6.8960$ & $1.0$ & $10^{4}$ \\
$1.8871 \times 10^{-1}$ & $1.7449 \times 10^{-3}$ & $7.3952$ & $1.4$ & $10^{4}$ \\
$2.0460 \times 10^{-1}$ & $3.0713 \times 10^{-3}$ & $7.4904$ & $1.8$ & $10^{4}$ \\
$2.1296 \times 10^{-1}$ & $4.3986 \times 10^{-3}$ & $7.5344$ & $2.2$ & $10^{4}$ \\
$2.1778 \times 10^{-1}$ & $5.6004 \times 10^{-3}$ & $7.5548$ & $2.6$ & $10^{4}$ \\
$2.2092 \times 10^{-1}$ & $6.7252 \times 10^{-3}$ & $7.5637$ & $3.0$ & $10^{4}$ \\
$2.2319 \times 10^{-1}$ & $7.7893 \times 10^{-3}$ & $7.5700$ & $3.4$ & $10^{4}$ \\
\bottomrule
\end{tabular}
\caption{Image quality metrics (Contrast, Sharpness, Shannon Entropy) for varying $\alpha$ and $\lambda$.}
\label{tabela2}
\end{table}

\clearpage
\section{Conclusion}

We have presented a fractional variational filtering method for denoising Raman spectra and image noise. By formulating the problem in the frequency domain, we derive an analytical solution that balances peak position and area in one-dimensional signals and preserves the essential features of images.

The key contribution of this work is the use of the Fourier transform to efficiently solve the variational problem with a fractional derivative regularization term, which significantly reduces computational complexity compared to traditional methods that require the inversion of dense matrices.

The incorporation of fractional derivatives in the regularization term of the functional allows refined control over the smoothing process, overcoming the limitations of classical methods based on integer-order derivatives. In the one-dimensional case, we observe that increasing $\alpha$ allows smaller values of $\lambda$ to be considered.

A crucial direction for future research is the development of methods for the automatic and adaptive selection of $\alpha$ and $\lambda$. While the Shannon entropy method presented here is a step in this direction, more robust, data-driven approaches are needed to make the method more user-friendly and applicable to a wider range of experimental data without manual tuning.

Results from simulated data demonstrate the method's effectiveness in recovering Raman signals, maintaining critical parameters such as peak position, intensity, and area even in the presence of noise.

The generalization to 2D images, using the fractional Laplacian, highlights the versatility of the approach, showing superior performance in preserving edges and textures compared to local operators. Quantitative analysis with contrast, sharpness, and entropy metrics confirms that the parameter $\lambda$ acts as a smoothing agent, while $\alpha$ promotes detail enhancement, making the balanced choice of these parameters crucial to avoid loss of information or noise amplification.

Based on these results, we conclude that the proposed method offers a robust and flexible mathematical tool for signal and image processing, combining the advantages of fractional calculus, variational optimization, and the computational efficiency of the Fourier transform.

%


\end{document}